\documentclass[a4paper,11pt]{article}
\usepackage{pos,color,soul}

\newcommand{\beq}{\begin{equation}}
\newcommand{\eeq}{\end{equation}}

\newcommand{\sd}{\mathrm d}

\newcommand{\om}{\omega}
\title{Estimating the thermal photon production rate \\ using lattice QCD}
\ShortTitle{Estimating the thermal photon production rate using lattice QCD}
\author*[a]{Csaba T\"or\"ok}
\author[b]{Marco C{\`e}}
\author[c]{Tim Harris}
\author[a]{Ardit Krasniqi}
\author[a,d,e]{Harvey B. Meyer}
\author[a]{Arianna Toniato}

\affiliation[a]{$PRISMA^+$ Cluster of Excellence \& Institut für Kernphysik, Johannes Gutenberg-Universität Mainz,\\
  Saarstr. 21, 55122 Mainz, Germany}

\affiliation[b]{Theoretical Physics Department, CERN,\\
CH-1211, Geneva 23, Switzerland}

\affiliation[c]{School of Physics and Astronomy, University of Edinburgh,\\
EH9 3JZ, United Kingdom}

\affiliation[d]{Helmholtz Institut Mainz, Johannes Gutenberg-Universität Mainz,
Saarstr. 21, 55122 Mainz, Germany}

\affiliation[e]{GSI Helmholtzzentrum f\"ur Schwerionenforschung,
Planckstrasse 1, 64291, Darmstadt, Germany}

\emailAdd{ctoeroek@uni-mainz.de}
\abstract
{%{{{
We present results for the photon emission rate determined from the transverse 
channel vector correlator at fixed spatial momentum using two flavors of dynamical 
Wilson fermions at $T\sim$250~MeV. We estimate the transverse channel spectral 
function using the continuum extrapolated correlator by applying various fit 
ansätze with a smooth matching to the NLO perturbative result. We confront our 
estimate based on this channel with the latest results of our collaboration 
based on the difference of the transverse and longitudinal channels.

\vspace{1cm} \hspace{10.4cm}
CERN-TH-2021-171
}%}}}

\FullConference{%
 The 38th International Symposium on Lattice Field Theory, LATTICE2021
  26th-30th July, 2021
  Zoom/Gather@Massachusetts Institute of Technology
}

\begin{document}
\maketitle

\section{Introduction}
%{{{
Characterizing the properties of the quark-gluon plasma (QGP) is an ongoing effort 
both experimentally and theoretically.
From the experimental side, electromagnetic radiation has been considered as
a clean probe of the QGP, because the interaction between the plasma and the
photons is very weak due to the colorless nature of the latter.
However, experimental difficulties arise because the photons and dileptons 
can be created via different mechanisms during the entire time evolution of
an ultrarelativistic heavy ion collision and unravelling the different 
contributions is challenging.
The largest contribution to the signal is coming from decay photons,
i.e. from the decay of the final state hadrons.
This contribution is subtracted from the total signal, and what remains is called
the direct photons.
A more detailed overview about the experimental situation can be found in Refs.~\cite{David:2019wpt,Gale:2021zlc}.

In Ref.~\cite{Gale:2021zlc}, moreover, an unresolved discrepancy between
the measured direct photon yield of the PHENIX collaboration (at BNL)~\cite{Adare:2014fwh}
and the theoretical result is discussed.
Results for the direct photon yield from two other collaborations
have appeared since then, the data from the ALICE collaboration
(at CERN)~\cite{Adam:2015lda}
-- using a different experimental setup -- are also somewhat larger than
the theoretical result, but the results of the STAR collaboration~\cite{STAR:2016use} 
(also at BNL) are in good agreement with the theory.
The various discrepancies (between the results reported by STAR and PHENIX
or between PHENIX and theory) have not been resolved yet.

It is worth mentioning that the direct photon excess is in the transverse
momentum range where the dominant contribution to direct photons comes
from thermal photons, which are originating from the QGP and have
a transverse momentum about a few hundred MeV - few GeV typically.
The theoretical thermal photon yield used for comparison has been obtained
assuming a weakly coupled plasma and a time evolution described by 
relativistic hydrodynamics.
Basically, the thermal photon contribution is calculated by integrating 
the thermal photon rate over the entire spacetime volume.

In this contribution, we present an estimate of the thermal photon production
rate using the transverse channel Euclidean correlator which has not been
investigated yet on the lattice at finite temperature in this context.

The thermal photon rate can be expressed in terms of the vector channel
spectral function, $\rho_V = -\rho^{\mu}_{\mu}$, as
\beq
    \frac{\sd \Gamma_\gamma(k)}{\sd^3 k} = \frac{\alpha_\mathrm{em}}{\pi^2} \, \frac{\rho_V(\om=k,k)}{4k} \, \frac{1}{e^{k/T}-1} + \mathcal{O}(\alpha_\mathrm{em}^2).
    \label{eq:photonrate}
\eeq
The spectral function of the electromagnetic current is defined as
\beq
    \rho_{\mu\nu}(\om, {\bf k}) = \int \sd^4 x \, e^{i(\om t - {\bf k x})} \langle [J^\mathrm{em}_{\mu}(x), J^\mathrm{em}_{\nu}(0)^\dag]\rangle.
\eeq

In Ref.~\cite{Brandt:2017vgl}, the following combination of the transverse and the
longitudinal channel has been put forward
\beq
	\rho(\om,k,\lambda) \equiv \left(\delta_{ij} - \frac{k_i k_j}{k^2}\right) \rho_{ij}
	+ \lambda \left( \frac{k_i k_j}{k^2} \rho_{ij} - \rho_{00} \right)
	= 2 \rho_T + \lambda \rho_L.
\eeq
When $\lambda = 1$, this combination is identical to the vector channel 
spectral function.
Choosing $\lambda = -2$, this combination gives the difference between
the transverse and longitudinal channels, which has been investigated in Refs.~\cite{Brandt:2017vgl,Brandt:2019shg,Ce:2020tmx}.
When $\lambda=0$, the longitudinal contribution vanishes, and $\rho(\om,k,\lambda)$
is identical to (two times) the transverse channel contribution.
These various channels, especially the ones corresponding to $\lambda=-2$ and
$\lambda=0$ have various advantageous properties~\cite{Brandt:2017vgl} 
which may help in overcoming the spectral reconstruction problem.
As regards the thermal photon emission rate, due to the Ward-identity
$\om^2 \rho_{00}(\om,k) = k_i k_j \rho_{ij}(\om, k)$,
the above defined combination, $\rho(\om,k,\lambda)$, is independent of $\lambda$
at light-like kinematics, therefore one may substitute $\rho_V$ in
Eq. (\ref{eq:photonrate}) by $\rho(\om = k, k, \lambda)$ using an 
arbitrary value of $\lambda$.
In this contribution, we present results using the transverse channel, i.e.
using $\lambda=0$.

%}}}
\section{Continuum extrapolation}
%{{{
We have generated four ensembles at the same temperature,
about $T = 250$ MeV in the high temperature phase.
We use $N_f=2$ clover-improved dynamical Wilson fermions and the Wilson gauge action.
The pion mass is around $m_\pi \approx 270$ MeV, and the lattice spacings
are in the range of 0.033...0.066 fm.
Instead of the electromagnetic current, we use the isovector vector current,
which amounts to neglecting the disconnected contributions~\cite{Ce:2020tmx}.
We measured the Euclidean correlators using both the local and the conserved
discretizations for the currents both at source and sink, resulting in four
different discretizations of the correlator.
We normalized the correlators by the static susceptibility to avoid the need
of renormalization.
The two mixed discretizations are not independent, they can be transformed 
into each other using time reflections.
Therefore, these have been averaged appropriately, resulting in a total of
three different discretizations.

These discretized correlators then have been used to perform a simultaneous
continuum extrapolation.
More precisely, we carried out fits also by using only a single discretization
as well as using multiple discretizations simultaneously and built histograms
using the Akaike Information Criterion (AIC) weights of each fit.
We also implemented multiplicative tree-level improvement of the lattice
data, and carried out fits using the improved data as well.
The median of the resulting histogram has been used as the continuum limit 
value in the later stages of the analysis.
%}}}
\section{Reconstruction by fitting and matching to perturbation theory}
%{{{
In this section, we introduce the strategy we followed for obtaining
spectral information from the Euclidean correlator.
The spectral decomposition formula,
\beq
	G_{\mu\nu}(x_0,k) = \int_0^\infty \, \frac{\sd \om}{2 \pi} \, 
    \frac{\cosh[\omega(\beta/2 - x_0)]}{\sinh(\omega\beta/2)} \,
    \rho_{\mu\nu}(\om, k)
\eeq
relates the Euclidean correlator, $G_{\mu\nu}(x_0,k)$, measured on the lattice
to our objective, the spectral function.
Inverting this relation to calculate $\rho_{\mu\nu}$ is a notoriously difficult,
ill-posed problem.
The method, with which we approached the question of extracting valuable
information from the correlator was motivated by the analysis presented
in Ref.~\cite{Ghiglieri:2016tvj}.
There, the authors exploited perturbation theory results in the high-frequency
regime and also to slightly constrain the fit ans\"atze for the infrared regime.
We also assume the validity of perturbation theory in the UV regime down to a chosen
value of frequency, $\om_0$, which is called the matching frequency,
and use the perturbative spectral function in the large $\om$ regime.
For the infrared part, we use various fit ans\"atze, $\rho_{\mathrm{fit}}(\om)$.
The full expression for the spectral function takes the form
\beq
\rho(\om) = {\rho_{\mathrm{fit}}(\om)} ( 1-\Theta(\om, \om_0, \Delta) )
		  + {\rho_{\mathrm{pert}}(\om)} \Theta(\om, \om_0, \Delta).
\eeq
To connect the IR and UV regimes smoothly, we use a smoothed
Heaviside-function,
\beq
    \Theta(\om, \om_0, \Delta) = (1 + \tanh[(\om-\om_0)/\Delta]) / 2.
\eeq
For the perturbative part, we use the recent NLO calculation~\cite{Jackson:2019mop}
complemented by the Landau-Pomeranchuk-Migdal resummation (LPM)
~\cite{Jackson:2019yao,Ghisoiu:2014mh} to cure the singularity at the light-cone,
\beq
	\rho_{\mathrm{pert}}(\om) = \rho_{\mathrm{NLO}}(\om) + \rho_{\mathrm{LPM}}(\om).
\eeq
Concerning the fit functions for the IR part, we chose simple odd polynomial
functional forms as well as a piecewise polynomial ansatz.
The latter has been motivated by the IR behavior of $\rho_{\mathrm{pert}}(\om)$ and 
consists of two odd polynomials, one for $\om < k$ and another for $\om > k$.
These are matched continuously at the light-cone.
The fit parameters have been determined by $\chi^2$-minimization.
For the fit procedure, in order to avoid singular behavior, we regularized the
covariance matrix by multiplying the off-diagonal elements by 0.95.
The $\chi^2$-minimization have been performed for each jackknife sample
of the continuum correlator to estimate the statistical error.

%}}}
\section{Mock analysis}
\begin{figure}[h]
    \includegraphics[scale=0.60]{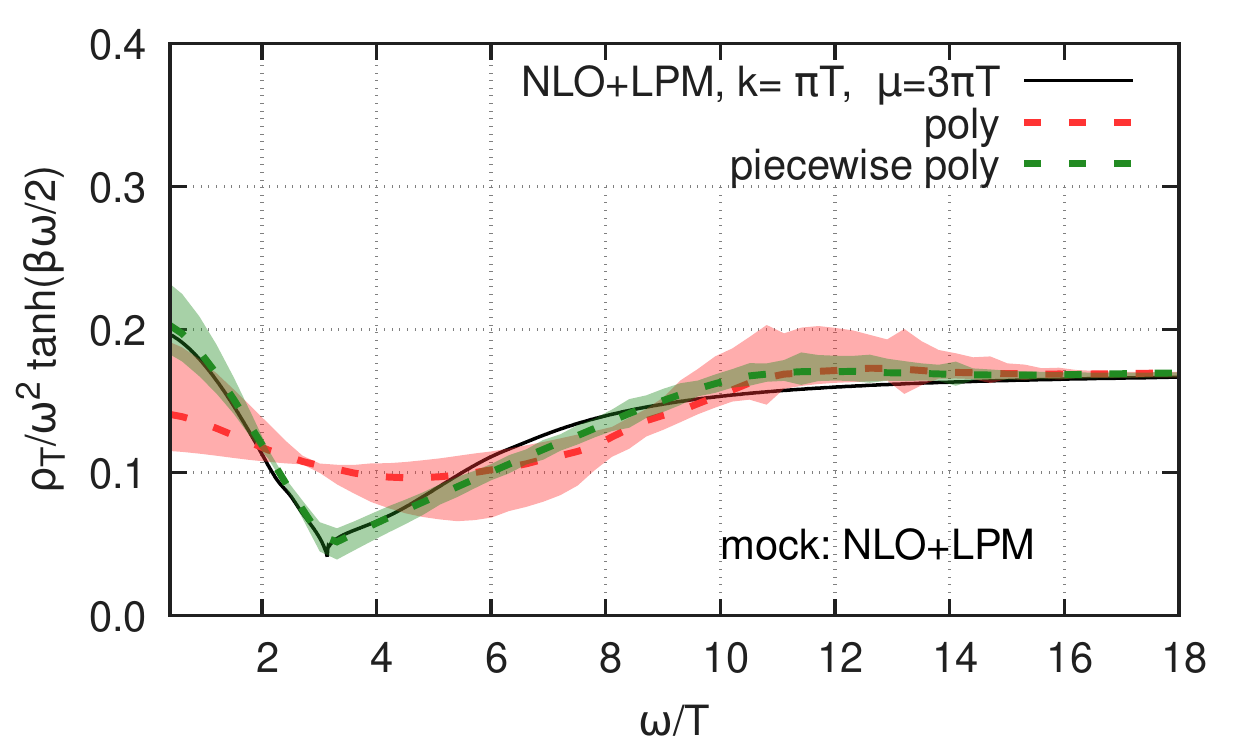}
    \includegraphics[scale=0.60]{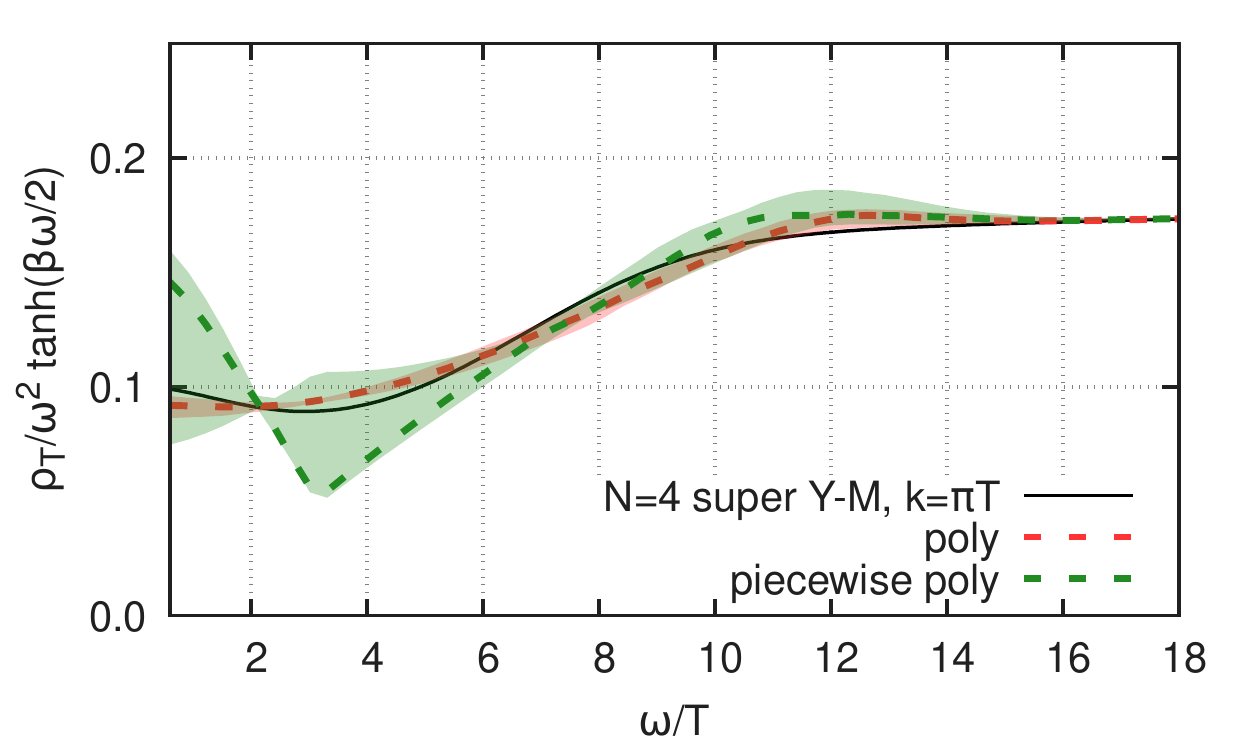}
\caption{Mock test results for the momentum $k/T = \pi$ using the 
	perturbative spectral function as input (left) or the spectral function
	of the $\mathcal{N}=4$ super Yang-Mills theory (right).}
	\label{fig:mock}
\end{figure}

%{{{
We have performed some mock tests using the perturbative spectral function,
$\rho_{\mathrm{pert}}$ as well as the spectral function of the $\mathcal{N}=4$ 
super Yang-Mills theory~\cite{Caron-Huot:2006pee} as input, mock spectral functions.
In these mock analyses we used a covariance matrix which has been rescaled
from the continuum covariance matrix in such a way that the relative errors
of the mock data are the same as those of the real lattice data.
This rescaled covariance matrix has not only been used for the
$\chi^2$-minimization, but also for the mock data generation.
With these mock analyses, we tested whether using a polynomial or a
piecewise polynomial ansatz can describe the input spectral functions well.
Representative examples of these mock tests are illustrated in Figure \ref{fig:mock}.

As one can observe in Figure \ref{fig:mock}, the piecewise polynomial
ansatz performs well in the case of the NLO+LPM spectral function,
but it underestimates the true value of the spectral function in the
case of the $\mathcal{N}=4$ super Yang-Mills theory, although the
statistical errors are large in that case.
The polynomial ansatz also returns spectral functions which are
in the same ballpark as the input one, although it cannot describe
well the dip at the light-cone in the case of the perturbative 
spectral function (Figure \ref{fig:mock}, left panel).
There, at $\om=k$, it significantly overestimates the true value.
%}}}
\section{Results from the lattice}
%{{{
When applying these fit ans\"atze for the real continuum extrapolated
lattice data, we made several variations of the parameters that may
influence the outcome of the $\chi^2$-minimization.
Regarding the perturbative input, we used two different renormalization
scales, $\mu = 2\pi T$ and $\mu = 3\pi T$.
The perturbative static susceptibility has been calculated according
to the formulae in Ref.~\cite{Vuorinen:2002ue}.
For an unknown non-perturbative coefficient in the corresponding expression,
we chose three values, one which has been originally estimated
in Ref.~\cite{Vuorinen:2002ue} and two other values which differ 
by around 20\% from this.
We used two different matching frequencies, $\om_0/T= $ 10 and 12.
For the width of the matching window, $\Theta(\om,\om_0,\Delta)$, we used three values,
$\Delta/T =$ 1.6, 2.0 and 2.4.
We used either two or three fit parameters, because we observed that allowing
for more fit parameters would result in overfitting in some cases.
The total number of available correlator data points was 9, but we
also performed $\chi^2$-minimization using only 8, 7 or 6 points.

\begin{figure}[h!]
	\includegraphics[scale=0.60]{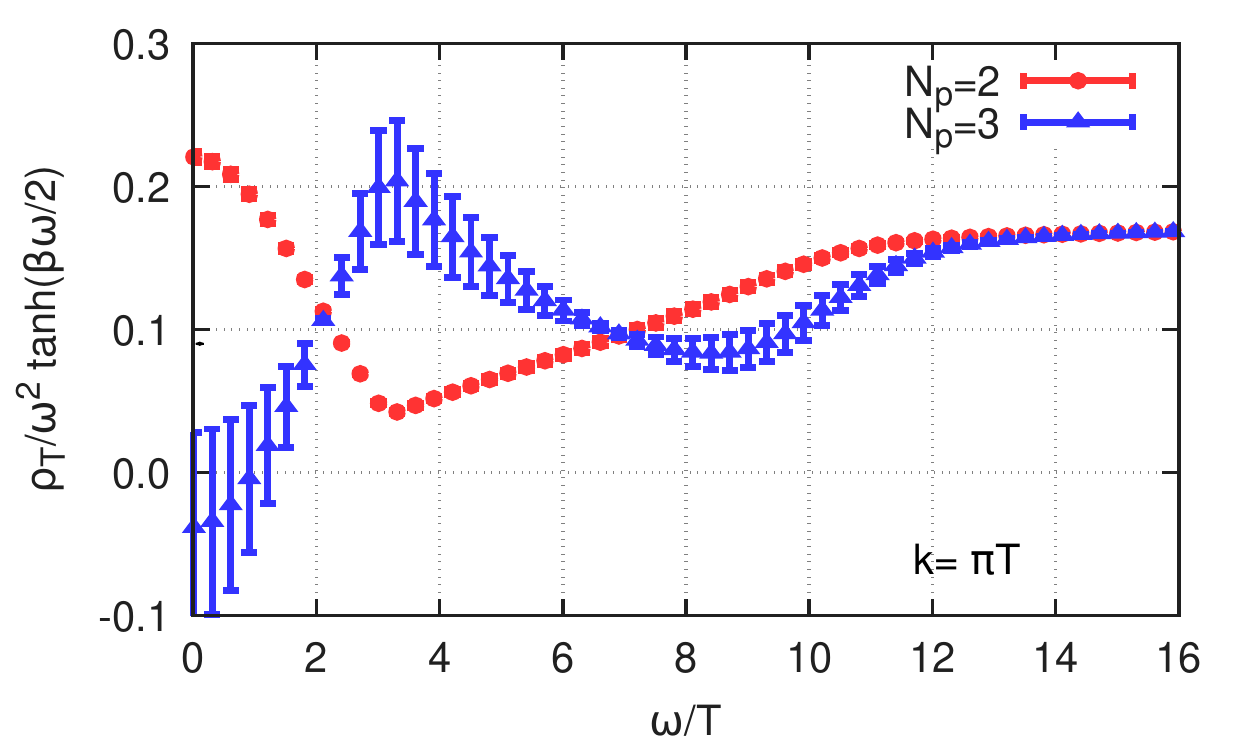}
    \includegraphics[scale=0.60]{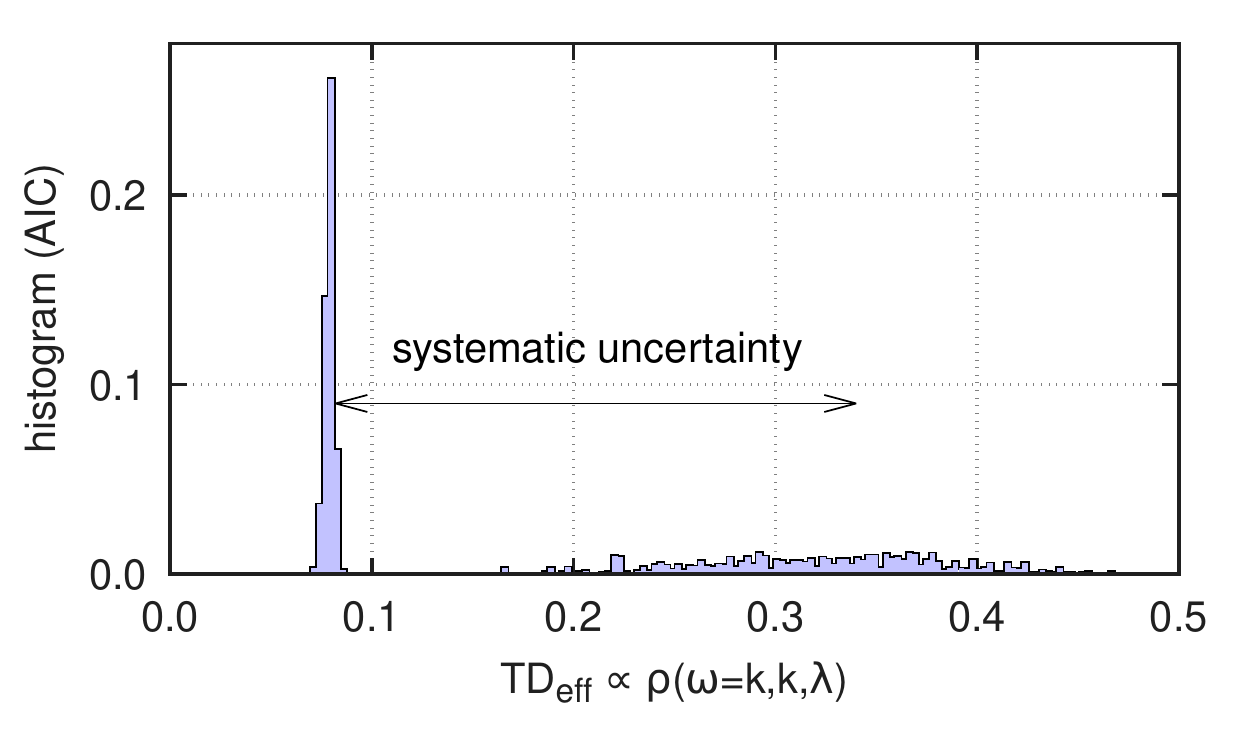}
	\caption{Left: Representative fit results using the piecewise polynomial
	ansatz at $k/T=\pi$. Right: Histogram of the effective diffusion 
	coefficient based on the AIC weights of the fits.}
	\label{fig:piecewise-poly_fits}
\end{figure}

The largest systematic uncertainty we observed was due to the different
fit forms.
A particular issue that arose when using the piecewise polynomial ansatz
is that we observed fits with acceptable $\chi^2/{ N_{\mathrm{dof}} }$ values
(also with acceptable $p$-values) which have either a maximum or a minimum at
the light-cone (Figure \ref{fig:piecewise-poly_fits}, left).
Fit results with local maximum at the light-cone were more frequent 
when we allowed three free parameters for the fit, but some fits 
with two parameters also produced similar results.
After assigning AIC weights to the various fits,
we built a histogram for the effective diffusion constant,
\beq
	D_{\mathrm{eff}}(k) = \frac{\rho(\om=k,k,\lambda)}{4 k} \frac{1}{\chi_s},
\eeq
which is the only non-kinematical factor in the expression of the thermal
photon rate (c.f. Eq. (\ref{eq:photonrate})).
Here, $\chi_s = G_{00}(x_0,{\bf 0})/T$ is the static susceptibility, for which
we used $\chi_s/T^2 = 0.880(9)_{\mathrm{stat}}(8)_{\mathrm{sys}}$~\cite{Ce:2020tmx}.
A histogram of the fit results for a particular momentum, $k/T=\pi$, can be seen
in the right panel of Figure \ref{fig:piecewise-poly_fits}.
Due to the large systematic uncertainty originating from the ambiguity
of the fit function, however, the histogram method for estimating the errors
of the effective diffusion constant is not applicable.
The systematic uncertainty can be estimated roughly as the difference between
the median values of the two disjoint histogram parts
(Fig. \ref{fig:piecewise-poly_fits}, right panel).

Although the systematic uncertainty can be estimated only roughly,
one can use the lattice results for $T D_{\mathrm{eff}}$ to provide
an upper bound on the thermal photon rate.
Since the piecewise polynomial ansatz can either have a dip-like minimum
or a spike-like maximum at light-like kinematics, it can be sensitive to
smaller as well as to larger values of the thermal photon rate.
We do not expect additional features in the infrared range at this
high temperature.

\begin{figure}[h!]
	\raisebox{0.65cm}
		{
		\includegraphics[scale=0.58]{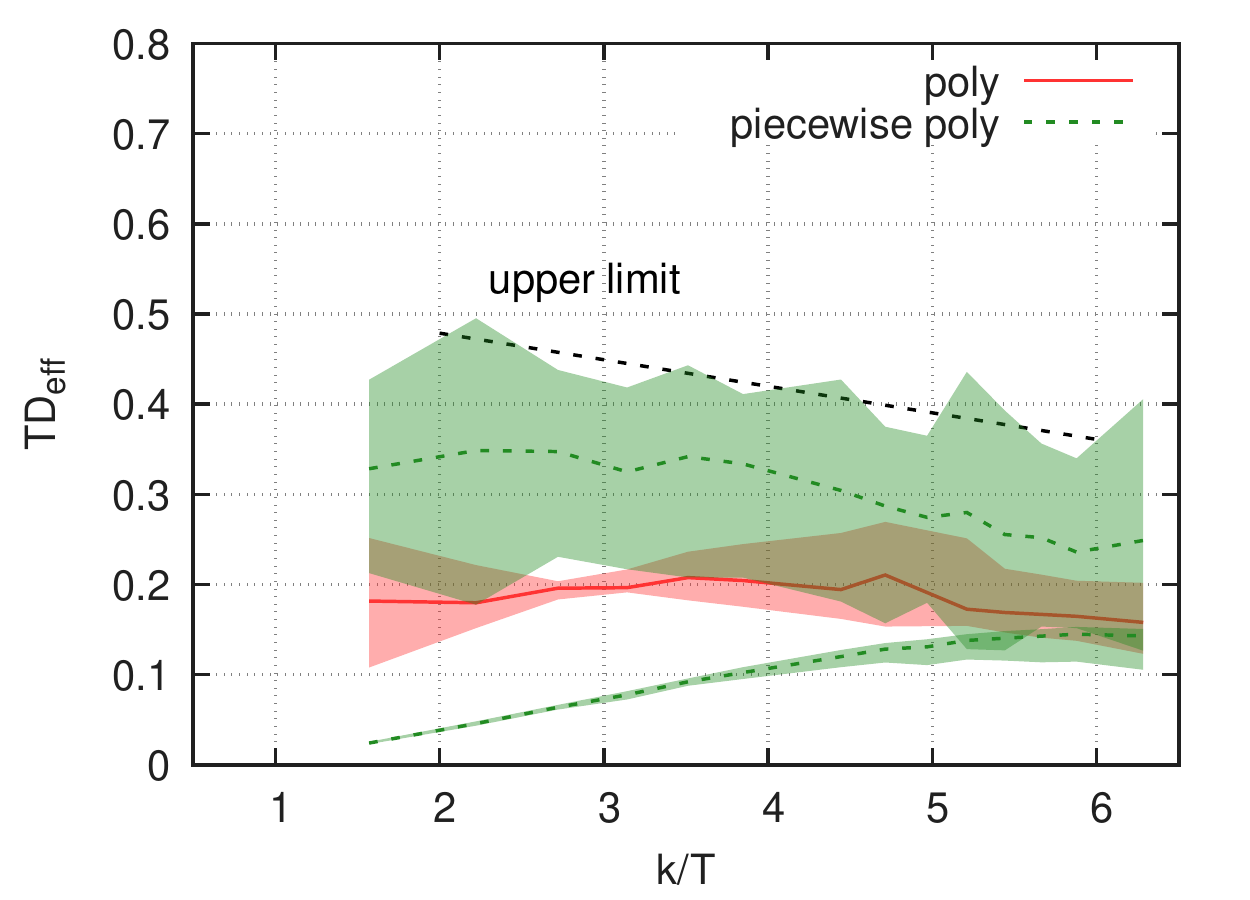}
		}
    \includegraphics[scale=0.32]{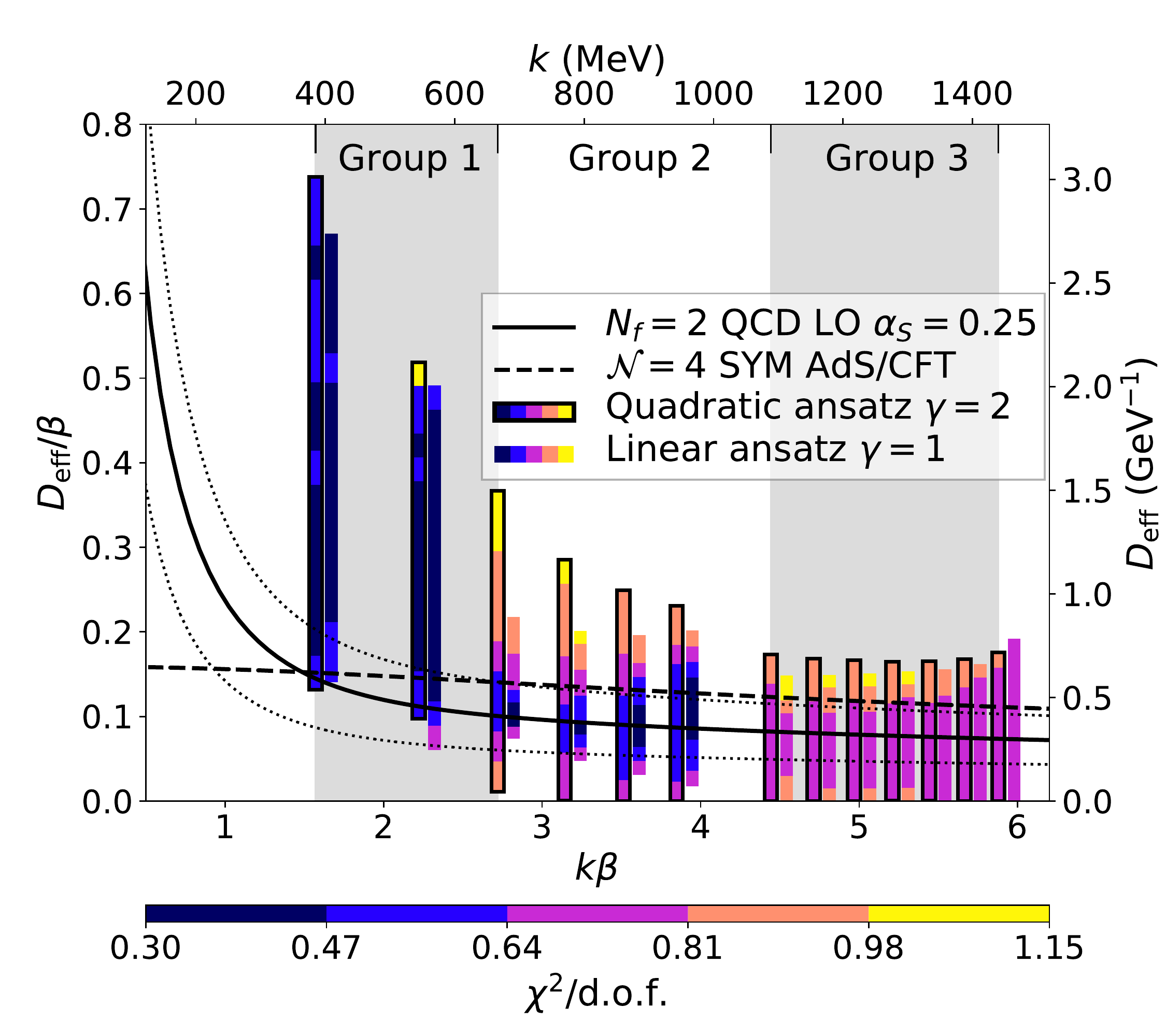}
	\caption
	{
	The effective diffusion constant, $TD_{\mathrm{eff}}$, obtained using
	the transverse channel correlator (left) or using the difference of 
	the transverse and longitudinal channels (right).
	On the right panel, the weak-coupling prediction as well as the result
	from the $\mathcal{N}=4$ super Yang-Mills theory are also included.
	}
	\label{fig:TDeff}
\end{figure}

The results for $T D_{\mathrm{eff}}$ from the transverse channel using
these ans\"atze are shown in Figure \ref{fig:TDeff}, left.
There, in the case of the piecewise polynomial ansatz, the central values of
the disjoint histogram parts are both illustrated with the dashed green line.
The systematic uncertainty illustrated in the right panel of
Figure \ref{fig:piecewise-poly_fits} is the distance of these two curves.
An upper bound estimate based on these fit results is included with a dashed
black line to guide the eye.

Thus, interpreting our approach to obtain an upper bound estimate on
the effective diffusion coefficient, we can compare our results obtained
by analyzing the transverse channel to the results obtained from 
analyzing the difference of the transverse and longitudinal channels,
performed earlier by our collaboration~\cite{Ce:2020tmx}, see 
Figure \ref{fig:TDeff}, right.
One can observe that the results are compatible with each other and
also with the perturbative result, though some of the transverse channel 
results are larger at $k/T > \pi$.
It is worth emphasizing that the analysis procedures were different in the
two cases.
For the analysis of the transverse channel, we performed fits for each
different momentum separately and also used the perturbative result
for the UV part, while for the T-L channel analysis, the available momenta
have been sorted into three groups and simultaneous multiple momentum
fits have been performed on these.
More details on the T-L channel analysis can be found in Ref.~\cite{Ce:2020tmx}.
Since the analysis of the T-L channel clearly favours $TD_{\rm eff}$
to lie below 0.2 at $k/T \gtrsim 4$, requiring consistency of the two 
analyses suggests that the solutions with a local maximum of the transverse 
spectral function at the lightcone are disfavoured.

%}}}
\section{Conclusions}
%{{{

A non-perturbative calculation of the thermal photon emission rate may provide 
valuable input to resolve current discrepancies in direct photon yields observed
between various experimental collaborations and between some collaborations
and the theoretical prediction.
In this work, we presented an estimate for the upper bound of the thermal
photon emission rate using $N_f=2$ $O(a)$-improved dynamical Wilson fermions.
We used four ensembles at around 1.2 $T_c$ with lattice spacings in the 
range 0.033...0.066 fm.
We performed a simultaneous continuum extrapolation using three discretizations of 
Euclidean transverse correlators of the isovector vector currents.
By utilizing the two-loop thermal perturbative spectral function for the UV regime,
we performed fits for the IR part of the spectral function using the
spectral decomposition formula.
We assumed simple fit ans\"atze (polynomial and piecewise polynomial) for the 
infrared spectral function, which performed reasonably well in mock
analyses.
When analyzing the lattice data, we investigated various sources of systematics
that could potentially affect the fit results and found that the largest contribution
to the systematics comes from specifying the fit function.
A reliable estimate of the systematic errors is hindered by this fact.
However, we could estimate an upper bound for the effective diffusion constant
or equivalently for the thermal photon emission rate.
The obtained fit results are in agreement with
a previous lattice determination using a different channel as well as with
the weak-coupling predictions.

%}}}
\section{Acknowledgements}
%{{{
This work was supported by the European Research Council (ERC) under the European Union’s
Horizon 2020 research and innovation program through Grant Agreement No. 771971-SIMDAMA,
as well as by the Deutsche Forschungsgemeinschaft (DFG, German Research Foundation) 
through the Cluster of Excellence “Precision Physics, Fundamental Interactions 
and Structure of Matter” (PRISMA+ EXC 2118/1) funded by the DFG within the 
German Excellence strategy (Project ID 39083149).
The work of M.C. is supported by the European Union’s Horizon 2020 research and
innovation program under the Marie Skłodowska-Curie Grant Agreement No. 843134-multiQCD.
T.H. is supported by UK STFC CG ST/P000630/1.
The generation of gauge configurations as well as the computation of correlators 
was performed on the Clover and Himster2 platforms at Helmholtz-
Institut Mainz and on Mogon II at Johannes Gutenberg University Mainz.
The authors gratefully acknowledge the Gauss Centre for Supercomputing e.V. 
(www.gauss-centre.eu) for funding project IMAMOM by providing computing time 
through the John von Neumann Institute for Computing (NIC) on the GCS Supercomputer 
JUWELS ~\cite{JUWELS} at Jülich Supercomputing Centre (JSC).
Our programs use the QDP++ library~\cite{Edwards:2004sx} 
and deflated SAP+GCR solver from the openQCD package ~\cite{Luscher:2012av}.
We are grateful to our colleagues in the CLS initiative for sharing the gauge field 
configurations on which this work is partially based.
%}}}


\begin{thebibliography}{99}
%{{{
\bibitem{David:2019wpt}
Gabor David,
\textsl{Direct real photons in relativistic heavy ion collisions}
Rept. Prog. Phys. 83, 4, 046301 (2020)
[arXiv:nucl-ex/1907.08893]

\bibitem{Gale:2021zlc}
Gale, Charles and Paquet, Jean-Fran\c{c}ois and Schenke, Bj\"orn and Shen, Chun,
\textsl{Multi-messenger heavy-ion physics}
[arXiv:nucl-th/2106.11216]

\bibitem{Adare:2014fwh}
Adare, A. and others, for the PHENIX collaboration,
\textsl{Centrality dependence of low-momentum direct-photon production in Au$+$Au collisions at $\sqrt{ s_{_{NN}} }=200$ GeV}
Phys. Rev. C 91, 6, 064904 (2015)
[arXiv:nucl-ex/1405.3940]

\bibitem{Adam:2015lda}
Adam, Jaroslav and others, for the ALICE collaboration,
\textsl{Direct photon production in Pb-Pb collisions at $\sqrt{s_{\rm NN}} =$ 2.76 TeV}
Phys. Lett. B 754, 235--248 (2016)
[arXiv:nucl-ex/1509.07324]

\bibitem{STAR:2016use}
Adamczyk, L. and others, for the STAR collaboration,
\textsl{Direct virtual photon production in Au+Au collisions at $\sqrt{s_{NN}}$ = 200 GeV}
Phys. Lett. B 770, 451--458 (2017)
[arXiv:nucl-ex/1607.01447]

\bibitem{Brandt:2017vgl}
Brandt, Bastian B. and Francis, Anthony and Harris, Tim and Meyer, Harvey B. and Steinberg, Aman
\textsl{An estimate for the thermal photon rate from lattice QCD}
EPJ Web Conf. 175, 07044 (2018)
[arXiv:hep-lat/1710.07050]

\bibitem{Brandt:2019shg}
Brandt, Bastian B. and C\`e, Marco and Francis, Anthony and Harris, Tim and Meyer, Harvey B. and Steinberg, Aman and Toniato, Arianna,
\textsl{Lattice QCD estimate of the quark-gluon plasma photon emission rate}
PoS LATTICE2019, 225 (2019)
[arXiv:hep-lat/1912.00292]

\bibitem{Ce:2020tmx}
C\`e, Marco and Harris, Tim and Meyer, Harvey B. and Steinberg, Aman and Toniato, Arianna,
\textsl{Rate of photon production in the quark-gluon plasma from lattice QCD}
Phys. Rev. D 102, 9, 091501 (2020)
[arXiv:hep-lat/2001.03368]

\bibitem{Ghiglieri:2016tvj}
Ghiglieri, J. and Kaczmarek, O. and Laine, M. and Meyer, F.,
\textsl{Lattice constraints on the thermal photon rate}
Phys. Rev. D 94, 1, 016005 (2016)
[arXiv:hep-lat/1604.07544]

\bibitem{Jackson:2019mop}
Jackson, G.,
\textsl{Two-loop thermal spectral functions with general kinematics}
Phys. Rev. D 100, 11, 116019 (2019)
[arXiv:hep-ph/1910.07552]

\bibitem{Jackson:2019yao}
Jackson, G. and Laine, M.,
\textsl{Testing thermal photon and dilepton rates}
JHEP 11, 144 (2019)
[arXiv:hep-ph/1910.09567]

\bibitem{Ghisoiu:2014mh}
Ghisoiu, I. and Laine, M.,
\textsl{Interpolation of hard and soft dilepton rates}
JHEP 10, 083 (2014)
[arXiv:hep-ph/1407.7955]

\bibitem{Vuorinen:2002ue}
Vuorinen, A.,
\textsl{Quark number susceptibilities of hot QCD up to g**6 ln g}
Phys. Rev. D 67, 074032 (2003)
[arXiv:hep-ph/0212283]

\bibitem{JUWELS}
Jülich Supercomputing Centre,
\textsl{JUWELS: Modular Tier-0/1 Supercomputer at the Jülich Supercomputing Centre}
Journal of large-scale research facilities 5, no. A135 (2019)

\bibitem{Edwards:2004sx}
Edwards, Robert G. and Joo, Balint,
\textsl{The Chroma software system for lattice QCD}
Nucl. Phys. B Proc. Suppl. 140, 832 (2005)
[arXiv:hep-lat/0409003]

\bibitem{Luscher:2012av}
Luscher, Martin and Schaefer, Stefan,
\textsl{Lattice QCD with open boundary conditions and twisted-mass reweighting}
Comput. Phys. Commun. 184, 519--528 (2013)
[arXiv:hep-lat/1206.2898]

\bibitem{Caron-Huot:2006pee}
Caron-Huot, Simon and Kovtun, Pavel and Moore, Guy D. and Starinets, Andrei and Yaffe, Laurence G.,
\textsl{Photon and dilepton production in supersymmetric Yang-Mills plasma}
JHEP 12 (2006) 015
[arXiv:hep-th/0607237]

%}}}
\end{thebibliography}
\end{document}